\documentclass[
preprint,
nobibnotes,
superscriptaddress,
amsmath, amssymb,
showkeys,
iop,
rpp,
12pt,
oneside
]{revtex4-2}

\usepackage{graphicx}
\usepackage{hyperref}
\usepackage{xcolor}
\usepackage{makecell}
\usepackage{forloop}

\graphicspath{{./figs/}} % the path of the figures folder
\begin{document}

\title{Anomalous Diffusion and Emergent Universality in Coupled Memory-Driven Systems}

\author{Nick Dashti}
\email{dashti.nick@gmail.com}
\affiliation{Australian Institute of Bioengineering and Nanotechnology, The University of Queensland, Brisbane, QLD, 4072, Australia}
\affiliation{School of Chemistry, University of Melbourne, Melbourne, Victoria 3010, Australia}

\author{M. N. Najafi}
\affiliation{Department of Physics, University of Mohaghegh Ardabili, P.O. Box 179, Ardabil, Iran}

\author{Debra J. Searles}
\email{d.bernhardt@uq.edu.au}
\affiliation{Australian Institute of Bioengineering and Nanotechnology, The University of Queensland, Brisbane, QLD, 4072, Australia}
\affiliation{School of Chemistry and Molecular Biosciences, The University of Queensland, Brisbane, QLD, 4072, Australia}
\affiliation{ARC Centre of Excellence for Green Electrochemical Transformation of Carbon Dioxide, The University of Queensland, Brisbane, QLD, 4072, Australia}
%==================================================

\begin{abstract}

Understanding how simple local interactions give rise to emergent exploration patterns is a fundamental question in statistical physics. We introduce a minimal model of two coupled agents that avoid retracing their own paths while being attracted to the trails left by one another. This system is inspired by, but not limited to, pheromone-guided insect navigation. The coupling of self-avoidance and attraction generates rich emergent behavior, including distinct anomalous diffusion regimes, non-Gaussian position distributions, and compressed exponential encounter statistics. Most notably, we identify new universality classes for coupled random walks, characterized by unique scaling laws and distributional properties that, to our knowledge, have not been previously reported. These findings advance the theoretical understanding of coupled stochastic processes with memory and interaction feedback, providing a framework for exploring transport phenomena in a broad range of multi-agent systems beyond biological contexts.

\end{abstract}

\maketitle

%==================================================
\section{Introduction}

Animal exploration strategies often deviate from simple random movement \cite{berg1993,codling2008}. This complexity arises particularly in species relying on chemical signals (semiochemicals) such as pheromones for communication and navigation \cite{shorey1973,wyatt2017}. 
Traditional random walk models have proven valuable for studying such processes, yet increasing evidence reveals the need for more sophisticated approaches to capture the nontrivial dynamics observed in nature \cite{dussutour2004}. Recent efforts have focused on deviations from classical diffusion processes, compelling the development of models that accommodate these complexities \cite{metzler2014,vilk2022}.

One such complexity arises in systems exhibiting anomalous diffusion, where the mean squared displacement (MSD) scales nonlinearly with time---a phenomenon ubiquitous in physical, financial, artificial, and biological systems \cite{balakrishnan1985,newman1999,bouchaud1990,avin2008,tsallis2009,bouchaud2003,van1992,voituriez2011}. In biological contexts, this behavior often stems from memory effects or interactions with evolving environments.
Insects, for example, modify their environment by laying down pheromone trails, which subsequently influence not only their own movements but also those of others in their vicinity \cite{deneubourg1990,sumpter2010,nakayama2023}.
This phenomenon extends to living cells that alter their surroundings by depositing biochemical signals or mechanically remodeling the extracellular matrix \cite{dalessandro2021,kranz2016,kranz2019}, as well as large animals that mark their territories \cite{giuggioli2011,potts2014}.
These feedback loops, reminiscent of self-organizing systems, lead to emergent behaviors that deviate sharply from classical diffusion models.
Non-Markovian random walks, in which movement decisions depend on the walker’s history, provide a framework for understanding these anomalous behaviors \cite{amit1983,ottinger1985,peliti1987,sapozhnikov1994,grassberger1996,pemantle2007,foster2009,metzler2014first,guerin2016,grassberger2017}.

One established approach for modeling such memory-driven dynamics is the true self-avoiding walk (TSAW), introduced by Amit, Parisi, and Peliti \cite{amit1983}. 
In the TSAW, the probability of stepping to site $i$, 
$p_i$, depends on the visitation history of each site: 
\begin{equation}
    p_i \propto e^{-\beta h_i},
\end{equation}
where $h_i$ represents the number of times site $i$ has been visited, and $\beta$ is a constant analogous to inverse temperature, controlling the strength of self-avoidance.
The TSAW extends the concept of the self-avoiding walk (SAW) \cite{Flory1953,deGennes1979} that is well-known for modeling the behavior of flexible chain polymers. Despite its name, the SAW is more accurately described as a `self-terminating' walk, where revisiting a site terminates the walk. In contrast, the TSAW employs a `softer' self-avoidance, where the walker tries to avoid retracing its steps.

Within the TSAW model, when self-interaction is repulsive ($\beta>0$), the time dependence of the MSD, $R^2$, as a function of time  exhibits distinct scaling laws in different dimensions, $d$, at large times \cite{amit1983,pietronero1983,obukhov1983,foster2009,grassberger2017,chebbah2022,dumaz2013,regnier2023uni,regnier2023rec,bremont2024}:
\begin{equation}
  \label{eq:tsaw}
  R^2 \sim 
  \begin{cases}
    t^{4/3} \quad & d=1\\
    t (\ln t)^{1/2} \quad & d=2 \\
    t \quad &d \ge 3.
  \end{cases}
\end{equation}
In contrast, introducing an attractive self-interaction ($\beta<0$) dramatically restricts the walker's movement. This leads to repetitive oscillations between two sites for any dimension $d$, resulting in a constant MSD at large times \cite{foster2009,chebbah2022}:
\[  R^2 \to C, \]
where $C$ is a constant depending on $d$ and $\beta$.
Introducing a saturation limit for $h_i$ (e.g., $h_i=0$ for unvisited site $i$, and $h_i=1$ if visited) transforms the self-attractive case, preventing walker trapping. This modified model, known as the self-attracting true random walk (SATW) or one-step reinforced random walk, \cite{prasad1996,foster2009,pemantle2007,ordemann2001,davis1990,agliari2012,chebbah2022}, exhibits a diverging MSD at large times:
\begin{equation}
  \label{eq:satw}
  R^2 \sim 
  \begin{cases}
    t \quad &d=1 \\
    t^{2/3} \quad &d=2 \\
    t \quad &d=3, |\beta| < |\beta_c| \\
    t^{1/2} \quad &d=3, |\beta| > |\beta_c| \\
  \end{cases}
\end{equation}
Here, $\beta_c$ marks a critical value in three dimensions, signifying the transition to a subdiffusive regime.

Natural environments present intricate scenarios, particularly when multiple interacting agents are involved \cite{alamgir2010,dashti2021}. 
A motivating example arises in the movement patterns of insects, where a trade-off between food exploration and mate-seeking influences collective dynamics. Insects rely on multiple semiochemicals, such as trail and sex pheromones, to guide their movements and interactions, resulting in complex, correlated behavior. Foraging insects tend to avoid revisiting previously explored locations, while the presence of sex pheromones attracts them toward areas marked by potential mates. This interplay of repulsion from self-deposited signals and attraction to cues from others creates a rich landscape of exploration strategies, with implications that extend beyond insect behavior to a broader class of systems where agents interact through modified environments.

This coupling between self-avoidance and mutual attraction gives rise to novel collective dynamics characterized by emergent scaling relations and new universality classes.
As far as we are aware, the exploration of such coupled agent systems---where memory and interaction feedback shape collective behavior---remains largely uncharted. This study advances the theoretical understanding of emergent scaling laws in coupled stochastic systems and highlights a rich area for further investigation.

In this work, our primary aim is to establish a clear and comprehensive understanding of coupled memory-driven dynamics in one dimension, where the simpler geometry allows for a more detailed exploration of observables and scaling relationships. 
Since two-dimensional systems present additional challenges and are of significant interest, we include key 2D results to demonstrate the broader relevance of our findings and to highlight key differences, which are presented in the following sections.

%==================================================
\section{The Model}

\begin{figure*}  \includegraphics[width=0.8\textwidth]{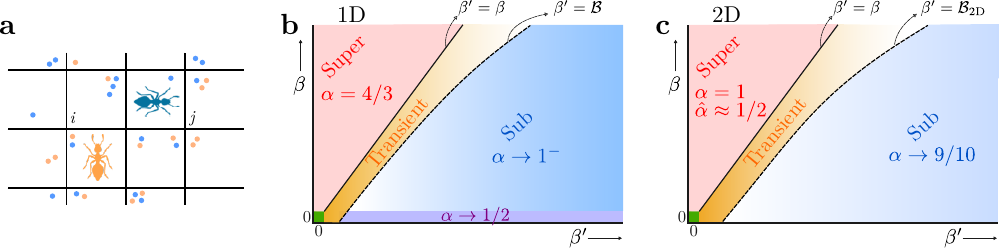}
  \caption{\label{fig:rws_phases} \textbf{Pheromonal random walks and phase diagrams.} 
  (a) Two insects (agents), A and B, represented in orange and blue respectively, collaborate to navigate in a stochastic and correlated manner, moving right, left, up, or down in two dimensions. As they visit sites, each insect deposits a unit of pheromone, indicated by orange circles for insect A and blue circles for insect B. In this configuration, the amount of the orange pheromone at the labeled site $i$ is $h_i^{(A)}=1$, while for the blue pheromone is $ h_i^{(B)}=3$ and at $j$ it is  $h_j^{(A)}=2, h_j^{(B)}=1$. The model assumes that each agent is more likely to move to sites with a greater amount of pheromone of the other agent and less of its own pheromone. 
  (b) The phase diagram for a one-dimensional system. We expect the scaling relation for MSD $R^2_t \sim t^{\alpha}$ at large times. When $\beta' \le \beta$ (red region), the system manifests superdiffusion with $\alpha = 4/3$. The dashed line represents the function $\beta'=\mathcal{B}(\beta)$, which separates the orange and blue regions and its exact form is unknown. For $\beta' > \mathcal{B}$ (blue region), the system exhibits subdiffusion with an exponent $\alpha$ approaching 1 from below ($\alpha \to 1^{-}$). When $\beta = 0$ and $\beta'$ is large (purple region), the system exhibits subdiffusion with $\alpha = 1/2$. The case where $\beta=\beta'=0$ corresponds to normal diffusion (the tiny green region).
  (c) The phase diagram for a two-dimensional system. The MSD follows the relation $R^2_t \sim t^{\alpha} (\ln t)^{\hat{\alpha}}$ at large times. Here, for $\beta \le \beta'$ (red region), $\alpha=1$ and $\hat{\alpha}=1/2$. For $\beta' > \mathcal{B}_{\text{2D}}$ (blue region), $\alpha\to 9/10$ and $\hat{\alpha}=0$.
  Unlike the 1D case, we do not observe a unique behavior for $\beta=0$.
  The orange regions, $\beta < \beta' < \mathcal{B}$, in (b) and (c) are transient regions where the system may not show stable behavior. 
  }
\end{figure*}

We consider a $d$-dimensional lattice over which two random walkers, A and B, undergo random walks. If we are modeling the paths of insects, A and B could represent a male and a female. The process starts when the walkers are $D_0$ apart. Each walker leaves a single, characteristic unit of `debris' upon stepping on the site $i$. Furthermore, in our model the agents prefer to avoid places they have been before, but are attracted to places that the other agent has been before.
Thus, this model can represent a pair of insects that are foraging and seeking each other with the debris being the pheromones from each insect. 
The amount (height) of pheromones released by the agent $X$ (A or B) at site $i$ up to time $t$ reads:
\begin{equation}
h_i^{(X)}(t) = \sum_{\tau=1}^{t} \delta_{x_{\tau}^{(X)}, i}, \quad h_i^{(X)}(0)=0 \; \forall i, \forall X
\end{equation}
where $x_{\tau}^{(X)}$ is the position of the agent $X$ at time $\tau$ and $\delta$ is the Kronecker delta.  In the example of the insects, this measures the amount of pheromone left by each of the insects at each site as a function of time (see Fig. \ref{fig:rws_phases}a). Since the same amount of pheromone is left each visit, $h_i$ also quantifies the number of visits to each site.
Each agent engages with its own debris and that of the other agent differently. The crucial aspect of our model is the transition probability for the agent $X$ moving from the site $i$ to $j$, which is expressed as follows:
\begin{equation}
  \begin{split}
    p^{(X)}_{i \to j} &= \frac{e^{-\beta h_j^{(X)}} e^{+\beta' h_j^{(X')}} }{Z} \qquad \beta, \beta' \ge 0, \\
    Z &\equiv \sum_{j \in \text{nn}_i } e^{-\beta h_j^{(X)}} e^{+\beta' h_j^{(X')}}
  \end{split}
  \label{eq:p}
\end{equation}
It accounts for both the self-avoidance of agent $X$ and the attraction to the pheromone field of the other agent $X'$, reflecting the coupled nature of the interactions. Here, $Z$ is the normalization constant that ensures the transition probabilities from site $i$ to all its nearest neighbors sum to one. 
The term $\text{nn}_i$ represents the set of nearest neighbors of site $i$. 
A detailed pseudo-code of the simulation algorithm is provided in the Supplementary Material.

The coefficients $\beta \ge 0$ and $\beta' \ge 0$ serve as the avoiding and attracting factors, respectively. Note that this model does not allow the agents to stay at the same site at consecutive times, but the agents could move together to a neighboring site.  
According to the equation above, at each time step, an agent is more likely to move to sites containing a greater amount of the other agent's debris and less likely to move to sites with a greater amount of its own.
Our model assumes reciprocal behavior: the agents operate under the same transition rules. These rules incorporate memory effects, as the transition probabilities depend on the values of $h_i$, which are themselves determined by the agents' past locations.

%==================================================
% \subsection{Measures}

We are interested in calculating the mean-squared distance between two agents, A and B:
\begin{equation}
D^2_t \equiv \left\langle ( x^{(A)}_t - x^{(B)}_t )^2 \right\rangle,
\end{equation}
and the mean end-to-end distance of the path of a single agent, equivalent to MSD for each agent:
\begin{equation}
R^2_t \equiv \left\langle \left( x_t - x_0 \right)^2 \right\rangle,
\end{equation}
where $\langle \cdot \rangle$ denotes the ensemble average.  
The observable $D^2_t$ captures how the relative positions of the two agents evolve over time and serves as a direct measure of interaction range and coordination. 
Since the model assumes reciprocal behavior, the average MSD will be the same for both agents. 

The scaling ansatz held for many anomalous diffusion processes is given by \cite{bouchaud1990,metzler2014,balakrishnan1985}:
\begin{equation}
  \label{eq:R}
  R^{2}_{t\to\infty} \sim t^\alpha (\ln t)^{\hat{\alpha}},
\end{equation}
where $\alpha=1$ and $\hat{\alpha} = 0$ for normal diffusion, i.e., for non-correlated standard random walks. 
Superdiffusion refers to the case $\alpha > 1$ or $\alpha=1,\hat{\alpha}>0$, while subdiffusion corresponds to $\alpha < 1$. 
Additionally, the increment statistics display a similar behavior to MSD at large times, offering another measure, called mean-squared increment (MSI) \cite{chebbah2022}:
\begin{equation}
  \label{eq:Delta}
  \Delta^2_{t,T} \equiv \left\langle \left( x_{T+t} - x_{T} \right)^2 \right\rangle
\end{equation}

To analyze the evolving interactions between agents, we track two key measures in the period up to time $t$: the number of encounters between two agents ($m$) and the total time they spend at the same location ($\mathcal{T}$). In the case of non-correlated normal random walks moving in one dimension, the probability distributions of these metrics, $P(m,t)$ and $P(\mathcal{T},t)$, exhibit Gaussian behavior. However, in two dimensions, these distributions are exponential. Moreover, the probability distribution of each agent's position, $P(x,t)$, is known to be Gaussian regardless of the dimensionality \cite{van1992}. 
Calculating the statistical measures for uncorrelated normal random walks is not the primary focus of this work. Therefore, further details are provided in the Supplementary Material.

The key difference between our model and previous ones lies in the source of attraction within Eq. \eqref{eq:p}. In our model, the attraction stems from the path laid down by the other walker (agent), not the walker's own path. This novel interaction gives rise to distinct universality classes and unique behaviors.  Furthermore, it provides a model for physical systems such as the pathways of insects driven by different pheromones, and could apply to a broad range of systems including active biological cells and growth of neural networks \cite{grueber2010}.

%==================================================

\section{Results}

%% Computational Details
All data was obtained from random walk simulations with transition probabilities given by Eq.~\eqref{eq:p} in 1D and 2D and were carried out using code developed in our group. We considered initial separations of $D_0=100$ for 1D and $D_0=10$ for 2D, with the self-avoidance coefficient set to $\beta = 0, 0.5$ or $1$ and various values of the attraction coefficient, $\beta'$. The ensemble averages were computed over 5 million samples for each parameter set, with some cases requiring over 10 million samples. During the simulation, measures such as the end-to-end distance, increment for each agent and the distance between the two agents were calculated at logarithmically spaced time intervals. Histograms of distances, positions, and encounter metrics were generated at specific time points. Output files were saved and stored for subsequent analysis to determine MSD, mean-squared distance, MSI, position distributions, number of encounters, total encounter durations, and their scaling behaviors.

Due to the reciprocal nature of our model, statistical measures such as $R^2_t$, and $P(x,t)$ are identical for both agents. Therefore, we report these measures only for the first walker. 
For the exponents, we report the nearest fractional values that accurately represent the numerical results.

The phase diagrams for the diffusion exponents in 1D and 2D (related to Eq.~\eqref{eq:R}) are shown in Fig.~\ref{fig:rws_phases}. In one dimension, we identify three distinct universality classes:
(i) For $\beta' \le \beta$, the diffusion exponent is $\alpha = 4/3$, characterizing a superdiffusion regime.
(ii) For $\beta' > \mathcal{B}$ (where $\mathcal{B}$ is a value of $\beta'$ that depends on $\beta$), the diffusion exponent approaches $\alpha = 1$ from below ($\alpha \to 1^{-}$), indicating a pseudonormal (and subdiffusion) regime. This behavior deviates from normal diffusion, where $\alpha = 1$. We will demonstrate that the position probability distribution in this subdiffusion regime is non-Gaussian. The intermediate region ($\beta < \beta' < \mathcal{B}$) exhibits a transition between the two regimes. Investigating this region poses practical challenges, as it may require extremely long simulations, and we skip it in this manuscript.
(iii) When $\beta = 0$, and $\beta'$ is large, the system represents subdiffusion with $\alpha = 1/2$. Importantly, all these results are independent of the initial distance between the two walkers, $D_0$.

\begin{figure*} \includegraphics[width=0.98\textwidth]{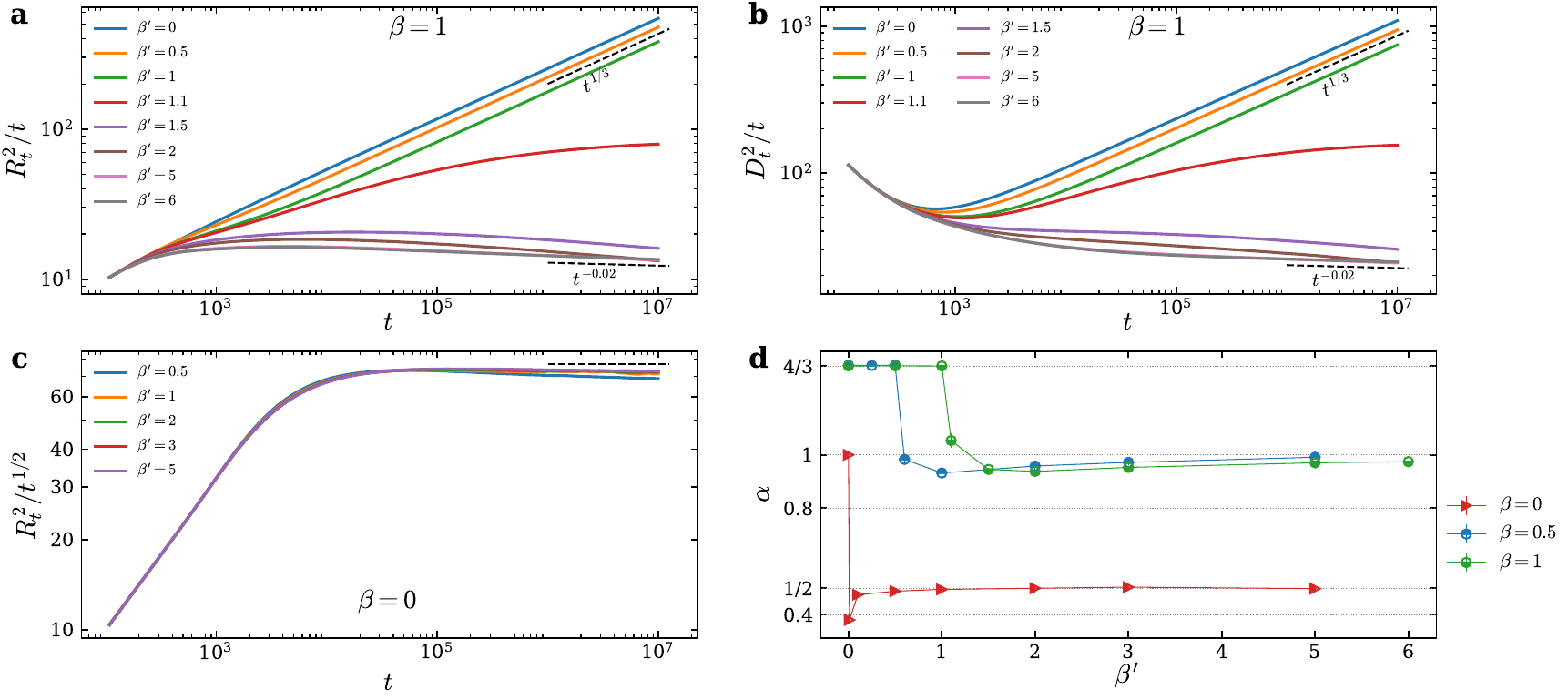}
  \caption{\label{fig:msd} \textbf{Mean-squared displacement and distance for a 1D model system.}  
  Here we plot the statistical measures for the first agent. Due to the reciprocal interaction scheme, the statistics for both agents are identical. (a) $R^2_t/t$ versus time $t$ is plotted for various $\beta'$ with fixed $\beta=1$. Superdiffusion ($\alpha=4/3$; $R_t^2/t \sim t^{1/3}$ for large $t$) is evident for $\beta'\le 1$, while subdiffusion ($\alpha \to 1^{-}; R_t^2/t \sim t^{0^-}$ for large $t$) occurs for $\beta' \ge 2$. 
  (b) The mean-squared distance between two agents, $D_t^2$ is depicted, demonstrating a similar asymptotic behavior to the mean-squared displacement of each agent.
  (c) $R^2_t/t^{1/2}$ is plotted versus $t$ for $\beta=0$ and $\beta'>0$, showing that for large values of $\beta'$, $\alpha=1/2$ ($R^2_t/t^{1/2} \to const $) at large $t$. 
  (d) The dependence of $\alpha$ versus $\beta'$ is plotted for various values of $\beta=0, 0.5$, and $1$. The error bars are smaller than symbol sizes.
  }
\end{figure*}

Figure~\ref{fig:msd}a demonstrates the time dependence of an agent's MSD for various values of $\beta'$, with a fixed coefficient $\beta = 1$ and $D_0 = 100$. Figure~\ref{fig:msd}b displays the mean-squared distance between two agents over time, again with varying $\beta'$ values. These curves exhibit similar asymptotic trends compared to $R^2_t$. The case of $\beta = 0$ is unique, as shown in Fig.~\ref{fig:msd}c: for large values of $\beta'$, the exponent $\alpha = 1/2$. Figure~\ref{fig:msd}d illustrates how the exponent $\alpha$ depends on $\beta'$ for three values of $\beta$ (0, 0.5, and 1).

\begin{figure*} 
  \includegraphics[width=0.98\textwidth]{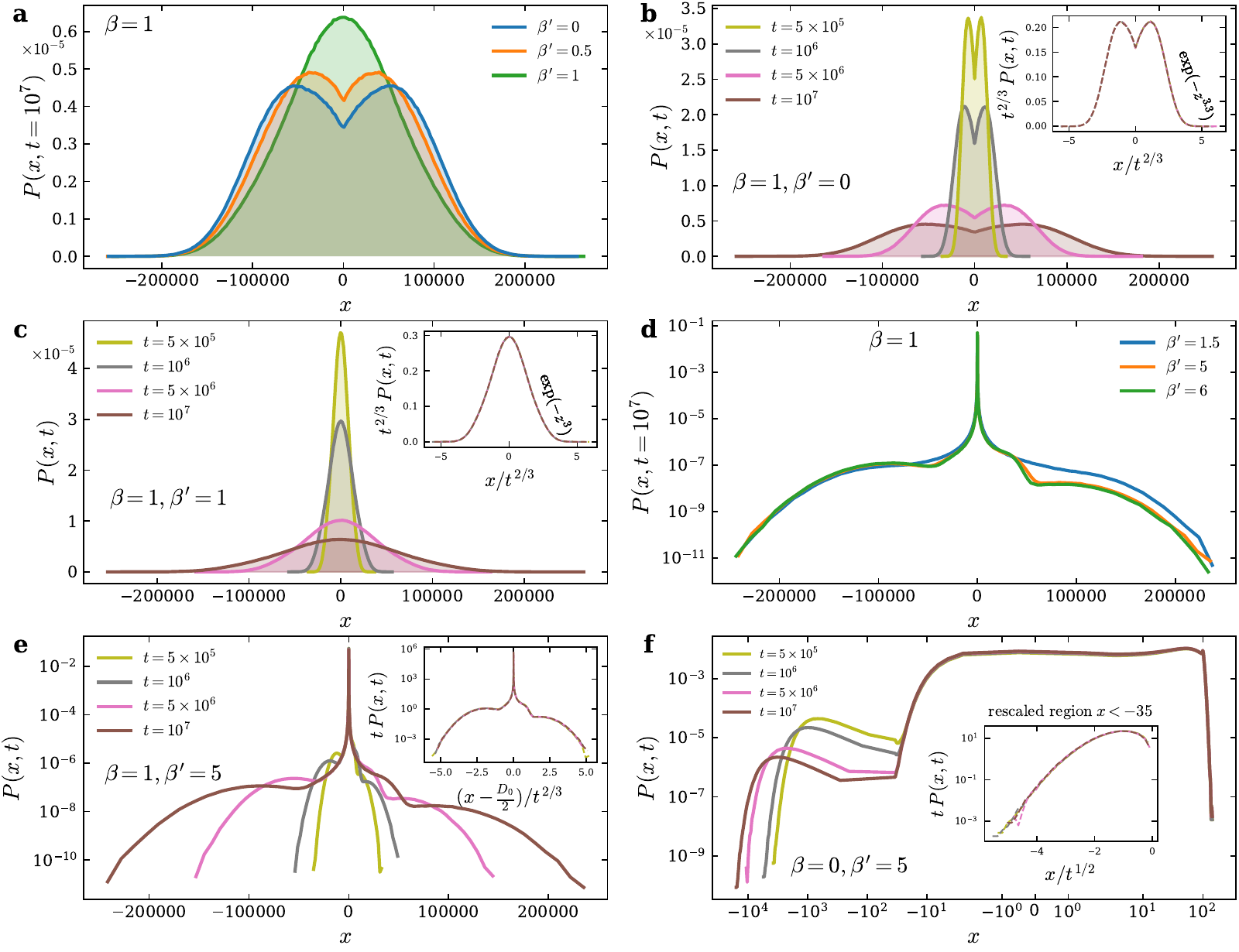}
  \caption{
    \label{fig:Px} 
    \textbf{Position probability distribution function.} 
    (a) The $P(x,t)$ of the first agent for $\beta'=0, 0.5$, and $1$ (fixed $\beta=1$) at time $t=10^7$. Distributions for $\beta'\le \beta$ are symmetric. With increasing $\beta'$, the bimodal peaks of $P(x,t)$ shift towards the origin, overlapping at $\beta'=1$.
    (b) and (c) Time evolution of the probability distribution function for $\beta'=0$ and $\beta'=1$, respectively. Inset demonstrates that rescaling $x \to t^{-2/3} x$ and $P \to t^{2/3} P$ collapses the curves for different times onto a single curve. The tail of the collapsed curve follows $\exp(-z^{3.3})$ for $\beta' = 0$ and $\exp(-z^3)$ for $\beta' = 1$, where $z \equiv x/t^{2/3}$.
    (d) For $\beta'\ge \mathcal{B}$, where $\mathcal{B} \approx 1.5$ for $\beta=1$, the probability distributions $P(x,t)$ are asymmetric.  
    (e) For $\beta=1, \beta'=5$ (actually for all $\beta' \ge \mathcal{B}$), rescaling $x \to t^{-2/3}(x-D_0/2)$ and $P \to t P$ collapses the $P(x,t)$ curves for different times. Inset demonstrates this collapse.
    (f) Position probability distribution for $\beta=0$, $\beta'=5$. The first agent is largely confined within a region less than about $100$ and explores primarily $x<0$. Inset: For $x \lessapprox -35$, rescaling $x \to t^{-1/2}x$ and $P \to t P$ collapses the curves. Note that the initial distance between two agents for all simulations is $D_0=100$. 
    The maximum uncertainty in the estimation of the scaling exponents across all figures is approximately 0.05.
  }
\end{figure*}

The distinct universality classes governing the system's dynamics are reflected in the behavior of the position probability distribution, $P(x,t)$, and the encounter-related probability distributions, $P(m,t)$ and $P(\mathcal{T},t)$. These distributions themselves display unique universality classes. Fig.~\ref{fig:Px}a shows the normalized probability distribution function of the first agent for cases $\beta'=0, 0.5$ and $1$ with fixed $\beta=1$. 
These distribution functions corresponding to $\beta'\le \beta$ all are symmetric. In the absence of attraction of agents ($\beta'=0$), the probability distribution $P(x,t)$ exhibits two peaks. As $\beta'$ increases, these peaks shift towards the origin, ultimately overlapping when $\beta'=1$.  
To demonstrate the evolution of $P(x,t)$, we calculated the probability distributions at $t=5\times 10^5, 10^6, 5\times 10^6$, and $10^7$. Fig.~\ref{fig:Px}b illustrates this for the case of $\beta'=0$. As shown in the inset, rescaling $x \to x / t^{2/3}$ and $P \to t^{2/3} P$ leads to a collapse of the curves onto a single curve. This curve exhibits a thin tail (faster-than-Gaussian decay) that follows the relation $\exp(-a z^{3.3})$, with $a$ dependent on $\beta$ and $\beta'$. The thin-tailed behavior is also observed for all $\beta' \le \beta$, as exemplified by the case of $\beta'=1$ in the inset of Fig.~\ref{fig:Px}c where the tail follows the $\exp(-a z^{3})$.

When $\beta' > \mathcal{B}$, the distribution functions $P(x,t)$ lose their symmetry, where $\mathcal{B}\approx 1.5$ for fixed $\beta=1$. If we set the initial distance between two agents to $D_0=0$, one could expect a symmetric function for $P(x,t)$. The asymmetric behavior is evident in Fig.~\ref{fig:Px}d for $\beta'=1.5,5$, and $6$ with fixed $\beta=1$. For all $\beta' > \mathcal{B}$, the appropriate transformation to collapse the $P(x,t)$ curves for different times is $x \to (x-\frac{D_0}{2})/t^{2/3}$ and $P \to t P$. The inset of Fig.~\ref{fig:Px}e demonstrates this collapse for the case $\beta'=5$.

The position probability distribution functions for cases with $\beta=0$ and $\beta'>\mathcal{B} \approx 0.5$ exhibit strikingly different behavior. As shown in Fig.~\ref{fig:Px}f, for $\beta=0$ and $\beta'=5$, the first agent (initially at the origin) remains largely confined within a region $x \lessapprox 100$. Note that within the region $x \lessapprox -35$, rescaling $x \to t^{-1/2}x$ and $P \to t P$ collapses the distribution curves for different times $t$ onto a single curve. 

\begin{figure*} 
  \includegraphics[width=0.98\textwidth]{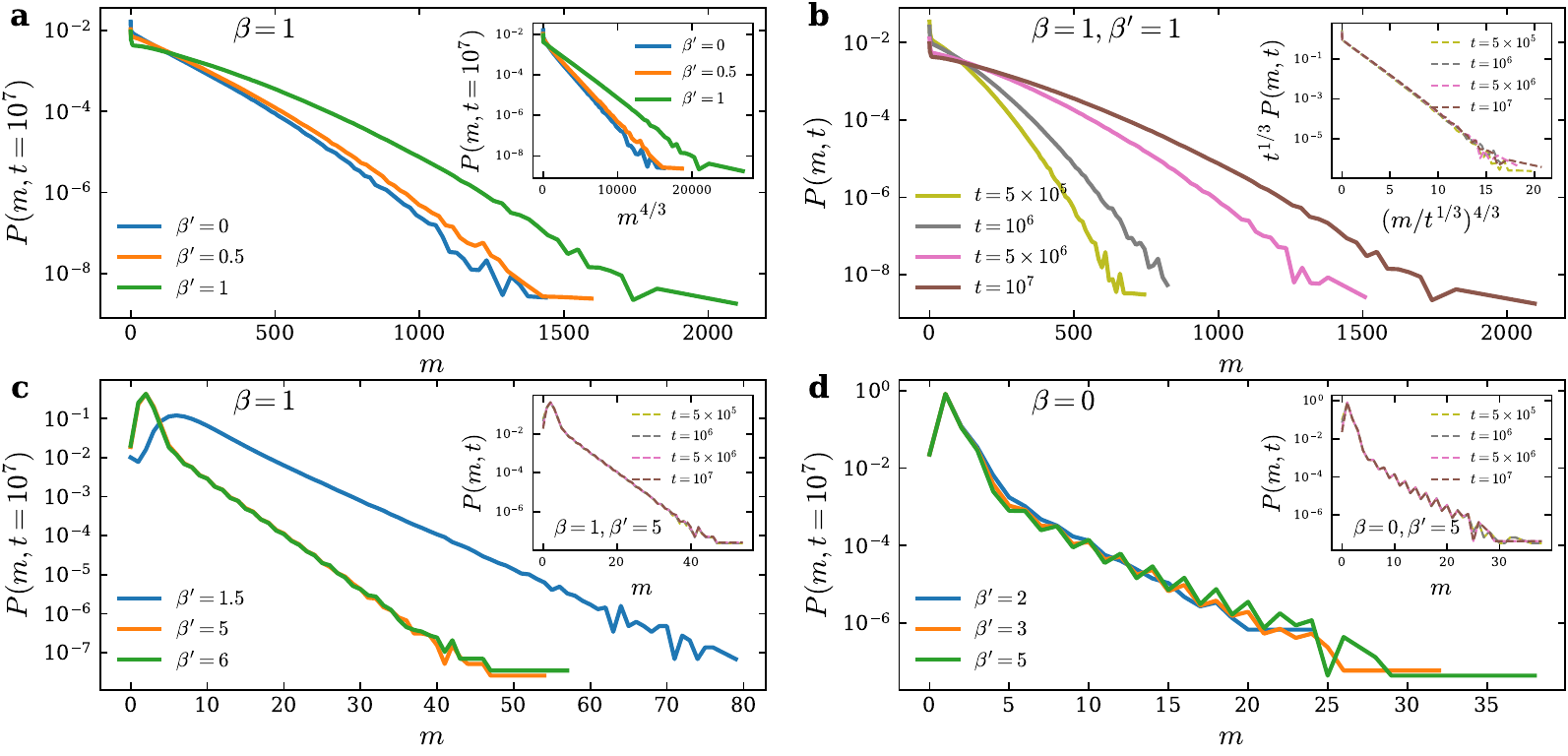}
  \caption{
    \label{fig:Pm} 
    \textbf{Probability distribution of encounters for a 1D model system.} 
    (a) Probability distribution $P(m,t=10^7)$ of encounters for  $\beta' = 0, 0.5$, and $1$ with fixed $\beta=1$ over the period $t=10^7$. The inset shows the distributions follow a compressed exponential function, $\exp(-az^{4/3})$, where $a = a(\beta, \beta')$.
    (b) Time-dependent probability distributions of encounters for $\beta=1, \beta'=1$. The inset demonstrates the collapse of curves onto a single curve using the transformation $m \to t^{-1/3}m$ and $P \to t^{1/3}P$. This scaling holds for all $\beta' \le \beta$.
    (c) $P(m,t=10^7)$ for $\beta'=1.5,5$, and $6$ with fixed $\beta=1$ in a semi-log plot. The distributions display exponential decay, $\exp(-am)$. The inset confirms that the distributions are time-independent.
    (d) $P(m,t=10^7)$ for case $\beta=0, \beta'=2,3,5$. Similar to (c), an exponential trend in the decay is observed, but with oscillations. The maximum uncertainty in the estimation of the scaling exponents across all figures is approximately 0.05.
  }
\end{figure*}

The distribution of the number of encounters between the agents ($m$) up to time $t$ is a crucial metric.  For insects that are foraging and seeking a mate, this would be the distribution of the frequency of the insects meeting.  Figure~\ref{fig:Pm}(a) shows the probability distributions $P(m,t=10^7)$ for $\beta' = 0, 0.5$, and $1$ in a semi-log plot. The inset highlights their compressed exponential behavior: $\exp(-az^{4/3})$, where $a$ depends on $\beta$ and $\beta'$. 
Figure~\ref{fig:Pm}b shows time-dependent distributions for $\beta=1, \beta'=1$. The inset demonstrates that by rescaling $m \to t^{-1/3}m$ and $P \to t^{1/3}P$, curves for different times collapse onto a single curve. This transformation holds for all $\beta' \le \beta$.
Figure~\ref{fig:Pm}c depicts the distribution functions for $\beta'=1.5, 3$, and $5$ (semi-log plot), revealing exponential decay: $\exp(-am)$. The inset confirms that these probability distributions are time-independent, so $P(m,t) \sim \exp(-am)$, with $a$ depending on $\beta$ and $\beta'$.
Figure~\ref{fig:Pm}d, for the case with self-interaction off ($\beta=0, \beta'>0$), exhibits a similar exponential behavior in  $P(m,t)$ as seen in Fig.~\ref{fig:Pm}c. Thus, the fundamental behavior of the exploration changes from superdiffusive to subdiffusive with the strength of the attract of the insects to each other's paths.

\begin{figure*} 
  \includegraphics[width=0.98\textwidth]{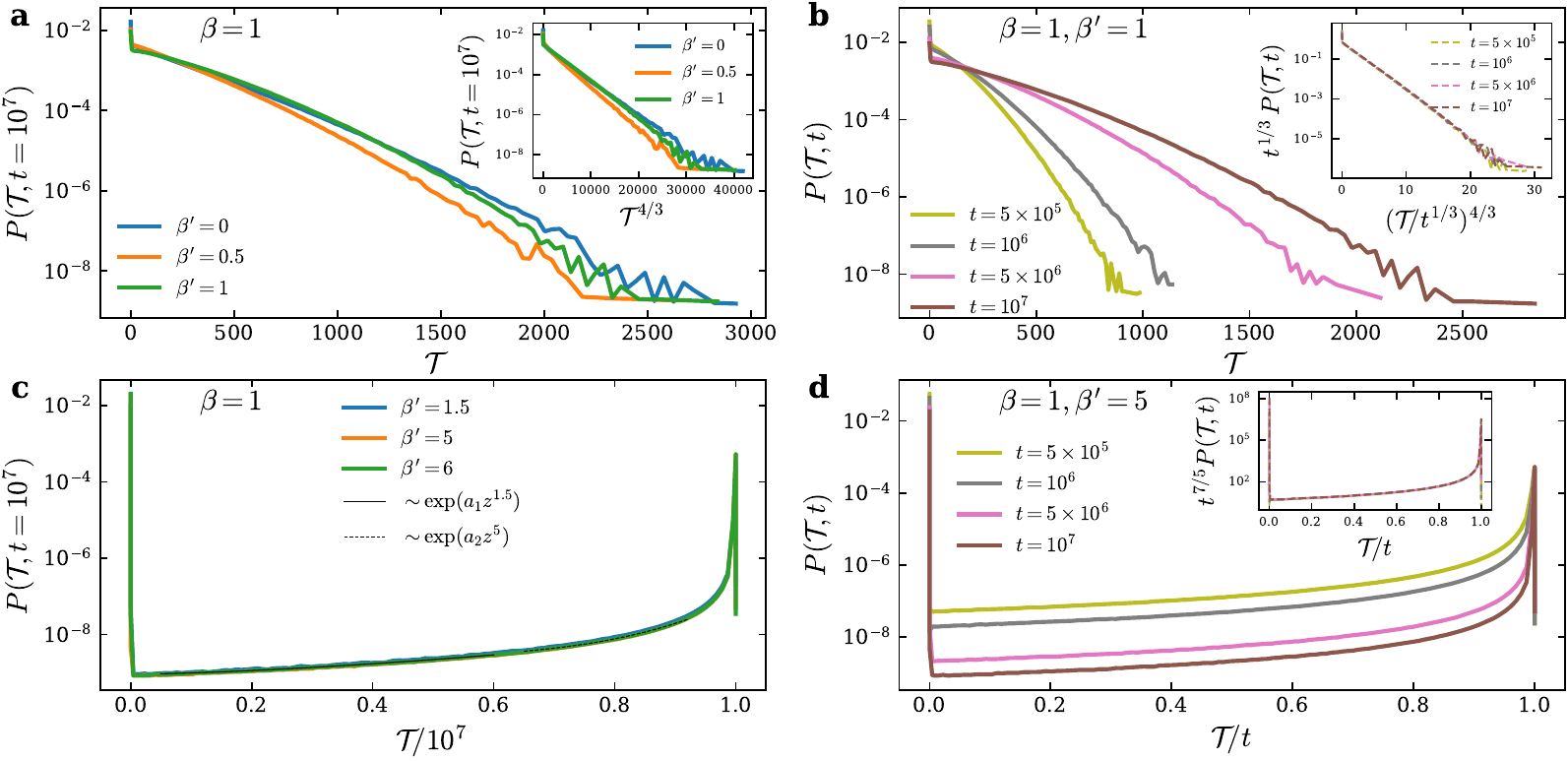}
  \caption{
    \label{fig:PT} 
    \textbf{Probability distribution of meeting duration for a 1D model system.} 
    (a) Probability distribution $P(\mathcal{T},t=10^7)$ of the total duration of meetings for  $\beta' = 0, 0.5$, and $1$ with fixed $\beta=1$. The inset shows the distributions follow a compressed exponential, $\exp(-az^{4/3})$, where the coefficient $a = a(\beta, \beta')$.
    (b) Time-dependent probability distributions for $\beta=1, \beta'=1$. The inset demonstrates that rescaling $\mathcal{T} \to \mathcal{T}/t^{1/3}$ and $P \to t^{1/3}P$ collapses the curves for different times onto a single curve.
    (c) $P(\mathcal{T},t)$ for superdiffusion region with $\beta'=1.5, 5$, and 6. Distributions exhibit high values for small $\mathcal{T}$, indicating limited duration of meetings. Two compressed exponential regimes are observed:  $\exp(a_1 z^{3/2})$ for $0.1 \lesssim \mathcal{T}/t \lesssim 0.6$ (black solid line) and $\exp(a_2 z^{5})$ for $0.6 \lesssim \mathcal{T}/t \lesssim 0.9$ (black dashed line). The coefficients $a_1$ and $a_2$ are functions of $\beta$ and $\beta'$.
    (d) $P(\mathcal{T},t =10^7)$ for $\beta=1, \beta'=5$. The inset shows that the curves for different times collapse under the rescaling $\mathcal{T} \to \mathcal{T}/t$ and $P \to t^{7/5} P$.
    The maximum uncertainty in the estimation of the scaling exponents across all figures is approximately 0.05.
  }
\end{figure*}

Another crucial metric is the probability distribution of the duration of meetings up to time $t$, denoted as $P(\mathcal{T}, t)$.  Fig.~\ref{fig:PT}a shows the probability distribution at $t=10^7$ for $\beta'=0, 0.5$, and 1 in a semi-log plot. The inset demonstrates that $P(\mathcal{T}, t)$ follows a compressed exponential function, $\exp(-az^{4/3})$, similar to the behavior observed for $P(m,t)$ in Fig.~\ref{fig:Pm}a.
Figure~\ref{fig:PT}b illustrates the time evolution of $P(\mathcal{T}, t)$ at various times ($t=5\times10^5, 10^6, 5\times 10^6$, and $10^7$). The inset shows that by rescaling $\mathcal{T} \to \mathcal{T}/t^{1/3}$ and $P \to t^{1/3}P$, these curves collapse onto a single curve. 
For $\beta'$ exceeding the transition value $\mathcal{B}$, the probability distributions exhibit markedly different behavior. As shown in Fig.~\ref{fig:PT}c, the probability distribution function has high values for finite $\mathcal{T}$, indicating that agents spend limited time together despite a higher attraction coefficient.  The three curves, corresponding to $\beta'=1.5, 5$, and 6, exhibit a compressed exponential behavior,  $\exp(a_1 z^{3/2})$, in the region $0.1 \lesssim \mathcal{T}/t \lesssim 0.6$, and $\exp(a_2 z^{5})$ for $0.6 \lesssim \mathcal{T}/t \lesssim 0.9$. The coefficients $a_1$ and $a_2$ depend on  $\beta$ and $\beta'$.
Fig.~\ref{fig:PT}d depicts the time-dependence of $P(\mathcal{T},t)$ for $\beta=1, \beta'=5$. The inset reveals that rescaling $\mathcal{T} \to \mathcal{T}/t$ and $P \to t^{7/5}P$ leads to a collapse of the curves.

\begin{figure*}  
  \includegraphics[width=0.98\textwidth]{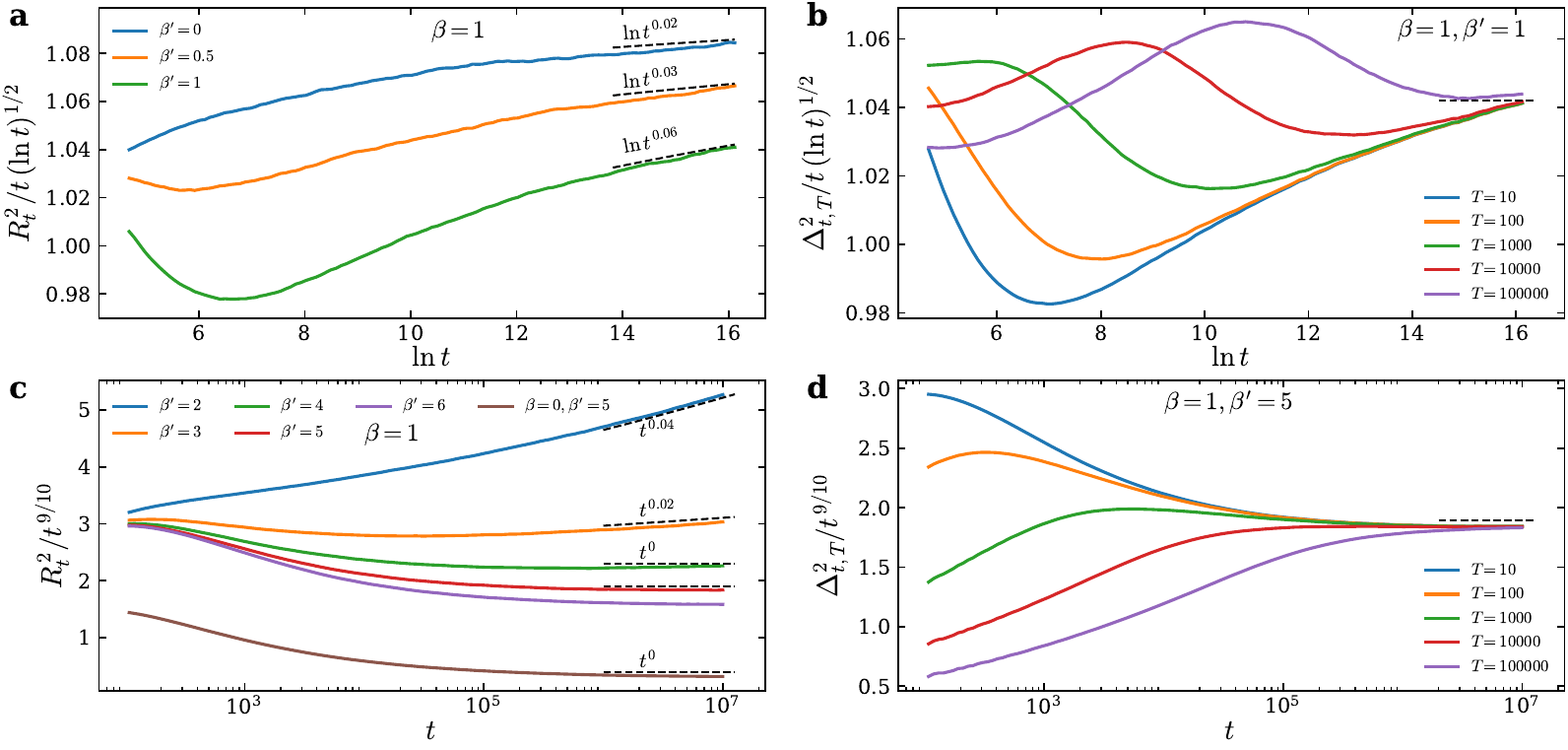}
  \caption{
    \label{fig:msd2} 
    \textbf{Mean-squared displacement for a 2D model system.}
    (a) MSD divided by $t (\ln t)^{1/2}$ versus $\ln t$  for $\beta'= 0, 0.5, 1$ with fixed $\beta=1$. While a slight deviation from the expected $\alpha = 1, \hat{\alpha}= 1/2$ behavior is observed for $\beta=1, \beta'=1$, further analysis is needed for confirmation. 
    (b) Mean-squared increment for various $T$ values, demonstrating a plateau at large $T$ and $t$ for $\beta=1, \beta'=1$.
    (c) MSD for various $\beta'=2, 3, 4, 5$, and $6$ with fixed $\beta=1$. The brown curve represents the case $\beta=0,\beta'=5$. The asymptotic behavior indicates a transition to a subdiffusion regime for $\beta' > \mathcal{B}_{\text{2D}}$, with  $\alpha = 9/10$ and $\hat{\alpha} = 0$. 
    (d) $\Delta^2_{t,T} / t^{0.9}$ versus time is plotted for various $T$ (fixed $\beta=1, \beta'=5$). The curves reaching a plateau confirm the exponents, $\alpha = 9/10$ and $\hat{\alpha} = 0$. 
  }
\end{figure*}

In two dimensions, the system exhibits markedly different behavior (see Fig.~\ref{fig:rws_phases}b). We initialize the simulation with the first agent (A) at the origin ($x=y=0$) and the second agent (B) at position  ($x=D_0, y=0$). The asymptotic behaviors do not depend on the value of $D_0$, at least for the rational, finite initial separations considered which were sufficiently large to avoid short-range effects but not so large as to introduce artificial decoupling. The results presented were obtained with $D_0=10$.
For $\beta' \le \beta$, superdiffusion is observed with $\alpha = 1$ and $\hat{\alpha} = 1/2$. While Fig.~\ref{fig:msd2}a suggests a slight deviation from this behavior for $\beta=1$ and $\beta'=1$, we turn to the mean-squared increment (Eq.~\ref{eq:Delta}) to obtain more reliable statistics. As shown in Fig.~\ref{fig:msd2}b, the curves for different values of $T$ ultimately reach a plateau for large $t$, confirming the superdiffusive regime.
When $\beta' \ge \mathcal{B}_{\text{2D}}$, the system transitions to a subdiffusion regime with $\alpha \to 9/10$ and $\hat{\alpha} = 0$, as illustrated in Figs.~\ref{fig:msd2}c-d. In our case, $\mathcal{B}_{\text{2D}} \approx 2$ when $\beta=1$.

\begin{figure*}  
  \includegraphics[width=0.95\textwidth]{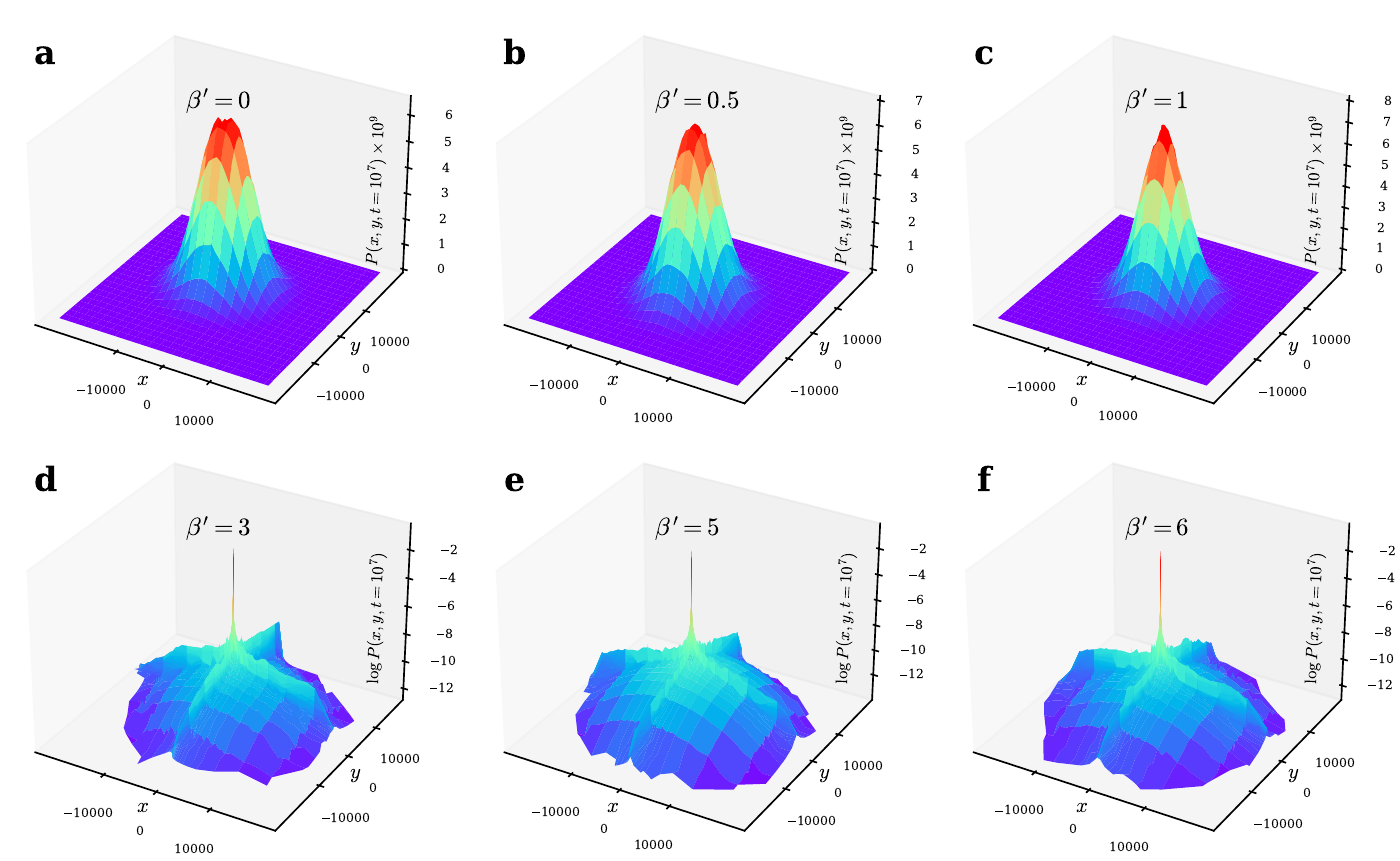}
  \caption{
    \label{fig:Px2} 
    \textbf{Position probability distribution function for a 2D model system.}
    (a)-(f) Position probability distributions at $t=10^7$ for $\beta'=0, 0.5, 1, 3, 4, 5$, with fixed $\beta=1$. (a)-(c) exhibit thin-tailed distributions while (d)-(f) show fat-tailed ones. 
  }
\end{figure*}

We analyze the position probability distribution of the first agent for various $\beta'$ with fixed $\beta=1$ (Fig.~\ref{fig:Px2}). For $\beta' \le \beta$, the distributions exhibit thin tails. However, when $\beta' > \mathcal{B_{\text{2D}}}$, the distributions transition to a fat-tailed (slower-than-Gaussian decay) behavior.

%==================================================
\section{Discussion}
The results of our 1D model reveal significant deviations from classical diffusion, driven by the trade-off between self-avoidance and attraction between agents. These dynamics manifest in distinct scaling behaviors, highlighting the complex interactions that arise from pheromone-guided exploration strategies.
Table~\ref{tab:1d} summarizes the key findings from our 1D model. The observed time-dependent probability distributions follow the scaling form:
\begin{equation}
  \label{eq:pu}
  P(u,t) \sim t^{-\zeta_u}f_u(u/t^{\nu_u}),
\end{equation}
where $u$ represents variables such as position ($x$ or $x-D_0$), number of encounters between two agents($m$), and total duration of encounters between two agents ($\mathcal{T}$). The exponents $\zeta_u$ and $\nu_u$ characterize the scaling behavior, and $f_u$ represents a universal scaling function. In thin-tailed distributions, the exponents $\zeta_u$ and $\nu_u$ are equal due to the normalizability of the probability distributions. However, in fat-tailed distributions, these exponents may not be equal \cite{cardy1996,kardar2007}.

The mean distance between two agents exhibits asymptotic behavior similar to MSD for all values of $\beta$ and $\beta'$ in any dimension (see Fig.~\ref{fig:msd}b and Figures S2 and S5 of the Supplementary Material). In a simplified scenario with independent walkers, we can envision the first agent as stationary at the origin while the second agent takes two steps at each time increment. This leads to a scaling equivalence: the distance between the second agent and the origin at time $t$ is analogous to the single-walker's MSD at time $2t$, $R^2_{2t}$.
When walkers' trajectories are correlated to each other's traces, a mean-field approach could be applicable. A crucial assumption for this approach is a uniform debris profile across the lattice. If we also maintain the reciprocal transition probability scheme, the simple interpretation of shifted origins for calculating the distance still remains valid.

For a purely self-interacting walker ($\beta > 0, \beta' = 0$), our model reduces to the well-studied true self-avoiding walk (TSAW) model \cite{amit1983,foster2009,chebbah2022}. 
For all values of $\beta' \le \beta$, we observe thin-tailed position distributions with $\zeta_x = \nu_x = 2/3$, but different shapes. Interestingly, increasing $\beta'$ from zero to $\beta$ transforms the distribution from a symmetric two-peaked form to a symmetric single-peak function (Fig.~\ref{fig:Px}a). Furthermore, the exponent $\alpha=4/3$, characteristic of the TSAW, emerges for all $\beta' \le \beta$ (Fig.~\ref{fig:msd}d). A mean-field approach could provide insights into this behavior: at large times, if we assume $h^{(A)}_i \approx h^{(B)}_i \approx \bar{h}$, the transition probability in Eq.~\eqref{eq:p} becomes proportional to $\exp[-(\beta-\beta')\bar{h}]$. Since $\beta' < \beta$, this effectively maps our model to the TSAW regime. 
However, for $\beta' = \beta$, the fluctuations of $h_i$s make this interpretation inapplicable. Moreover, for $\beta' > \mathcal{B}$, 
the strong sample-to-sample variability and the significant impact of rare events on the probability distributions of position and inter-agent distance make the mean-field method unsuitable.

A particularly intriguing aspect lies in the likelihood of meeting. Although Fig.~S6 of the Supplementary Material suggests Gaussian distributions for $m$ and $\mathcal{T}$ in non-correlated normal RWs, our model reveals compressed exponential functions for both $P(m,t)$ and $P(\mathcal{T},t)$, as shown in Figs.~\ref{fig:Pm}a-b, \ref{fig:PT}a-b. 
For $P(m,t)$ this slower decay compared to Gaussian distributions implies a higher frequency of observing a large number of meetings, and for $P(\mathcal{T},t)$ it implies a higher frequency of long total meeting durations within excursions. This is observed even though our walkers exhibit a greater increase in MSD with time than in non-correlated random walks. 
Thus, by tuning the values of $\beta$ and $\beta'$, we can optimize the distribution of the meetings. In this regime, $\zeta_m = \nu_m=1/3$ and $\zeta_{\mathcal{T}} = \nu_{\mathcal{T}}=1/3$.

Strong attraction to another agent through preferred movement to lattice points with their debris leads to the emergence of fundamentally distinct dynamics within the system.
For inter-attraction coefficients $\beta'$ exceeding both $\beta$ and a specific threshold $\mathcal{B}$, the exponent $\alpha$ asymptotically approaches $1^-$ (as seen in Fig.~\ref{fig:msd}a). While the near-unity exponent might suggest `normal' diffusion, the fat-tailed position probability distribution (Fig.~\ref{fig:Px}) reveals a significant departure from true Gaussian behavior. The term `pseudonormal' could aptly describe this regime.
Moreover, the meeting probabilities in this regime show non-Gaussian distributions. The probability $P(m,t)$ reaches a steady value (time-independent) for large times and also decays exponentially. 
The function $P(\mathcal{T},t)$ displays complex time-dependence (Figs.~\ref{fig:PT}c-d). While $P(\mathcal{T}, t)$ peaks near $\mathcal{T} \approx 0$, it also has a significant probability around $\mathcal{T} \approx t$. This indicates that while walkers typically have short encounters, there remains a possibility of longer durations. The scaling exponents in this regime are $\zeta_x = 1, \nu_x=2/3$, $\zeta_m = \nu_m=0$ and $\zeta_{\mathcal{T}} = 7/5, \nu_{\mathcal{T}}=1$.

A particularly intriguing regime emerges when self-repulsion is absent and strong inter-attraction exists between agents ($\beta' > \mathcal{B}$, $\beta=0$). In this regime, agents can become temporarily localized within regions, leading to subdiffusion with $\alpha=1/2$. Interestingly, this exponent matches the three-dimensional true self-avoiding walk (Eq.~\eqref{eq:satw}). 
The fat-tailed, asymmetric $P(x,t)$ distribution for the first agent (Fig.~\ref{fig:Px}f) reflects the initial separation distance $D_0$ and a typical directional bias: the first agent tends to move leftward, and the second agent rightward.

This regime exhibits strong sample-to-sample variability, where rare realizations---such as agents escaping localized regions and exploring large distances---significantly impact the ensemble-averaged statistics. Extensive sampling is required to capture these rare events accurately.
Although this behavior falls within the broader subdiffusion category, it contrasts sharply with the $\beta>0, \beta' > \mathcal{B}$ case, where agents remain more consistently localized. Notably, we do not observe this contrasting behavior in two or higher dimensions, likely due to the increased freedom of movement compared to the constraints imposed by one-dimensional open boundary conditions.
The scaling exponents for $P(x,t)$ are $\zeta_x=1, \nu_x=1/2$, but the scaling exponents for measures of encounter frequency and duration align with the pseudonormal regime ($\beta>0, \beta'>\mathcal{B}$).

\begin{table*}[b]
  \caption{\label{tab:1d}
  \textbf{The asymptotic behavior of various measures for 1D model systems.}
  All coefficients $a, a_1$ and $a_2$ for each measure are independent and may depend on the values $\beta$ and $\beta'$. $D_0$ is the initial ($t=0$) distance between two agents. $\mathcal{B}$ is a function of $\beta$.
  }
  \begin{tabular}{| c | c | c | c | c |}
    \hline 
    & 
    $0 \le \beta' \le \beta$& 
    $\beta' > \mathcal{B}, \beta > 0$ & 
    $\beta' > \mathcal{B}, \beta=0$ &
    $\beta' = \beta = 0$
    \\ \hline
    $R^2_{t\to \infty}$  & 
    $t^{4/3}$ & 
    $t^{1^-}$ & 
    $t^{1/2}$ &
    $t$
    \\ \hline
    $D^2_{t\to \infty} $ & 
    $R^2_{t\to\infty}$ & 
    $R^2_{t\to\infty}$ &
    $R^2_{t\to\infty}$ &
    $R^2_{t\to\infty}$
    \\ \hline
    $P(x,t) $ & 
    \makecell
    {
        $t^{-2/3} f \left( \dfrac{x}{t^{2/3}} \right)$, \\
        $f$ is thin-tailed
    } & 
    \makecell
    {
        $t^{-1} f \left( \dfrac{x-D_0/2}{t^{2/3}} \right)$ \\ 
        $f$ is fat-tailed
    } &
    \makecell
    {
        $t^{-1} f \left( \dfrac{x}{t^{1/2}} \right)$ for $x \ll 0$ \\
        $f$ is fat-tailed
    } &
    \makecell
    {
        $t^{-1/2} f \left( \dfrac{x}{t^{1/2}} \right)$, \\ 
        $f(z) = e^{-az^2}$ 
    } 
    \\ \hline
    $P(m,t) $ &
    \makecell
    {
        $t^{-1/3} f\left(\dfrac{m}{t^{1/3}} \right)$ \\
        $f(z) = e^{-a z^{4/3}}$
    } & 
    \makecell
    {
        % $f\left(m \right)$ \\
        $e^{-a m}$
    } &
    \makecell 
    {
        not conclusive, \\ 
        but should be \\ 
        similar to $\beta' > \mathcal{B}$, \\
        $\beta > 0$
    } &
    \makecell
    {
        $t^{-1/2} f \left( \dfrac{m}{t^{1/2}} \right)$, \\ 
        $f(z) = e^{-az^2}$ 
    }  
    \\ \hline
    $P(\mathcal{T},t) $ &
    \makecell 
    {
        $t^{-1/3} f\left(\dfrac{\mathcal{T}}{t^{1/3}} \right)$, \\ 
        $f(z) = e^{-a z^{4/3}}$
    } &
    \makecell
    {
        $t^{-7/5} f\left(\dfrac{\mathcal{T}}{t} \right)$ \\
        $f(z) = e^{a_1 z^{1.5}}$, \\ 
        $0.1 \lesssim z \lesssim 0.6$ \\
        $f(z) = e^{a_2 z^{5}}$, \\
        $0.6 \lesssim z \lesssim 0.9$
    } & 
    \makecell 
    {
        not conclusive, \\ 
        but should be \\ 
        similar to $\beta' > \mathcal{B}$, \\
        $\beta > 0$
    } &
    \makecell
    {
        $t^{-1/2} f \left( \dfrac{\mathcal{T}}{t^{1/2}} \right)$, \\ 
        $f(z) = e^{-az^2}$ 
    } 
    \\ \hline
  \end{tabular}
\end{table*}

Our simulations in two dimensions reveal two distinct regimes: superdiffusion and subdiffusion (Fig.~\ref{fig:msd2}, Table~\ref{tab:2d}). When $\beta' \le \beta$, we observe scaling exponents ($\alpha=1, \hat{\alpha}=1/2$) consistent with the true self-avoiding walk (TSAW). The probability distribution $P(x,y,t)$ exhibits symmetry and thin tails, further emphasized by $P(x,y=0,t)$ and $P(x=0,y,t)$ (see Fig. S10 of the Supplementary Material). Interestingly, two-dimensional scaling exponents deviate from the relation in Eq.~\eqref{eq:pu}, instead including logarithmic terms in the distribution functions for position and encounter-related measures:
\begin{equation}
  \label{eq:pu2d}
  P(u,t) \sim t^{-\zeta_u} (\ln t)^{-\hat{\zeta}_u} \, f_u \left( \frac{u}{t^{{\nu}_u} (\ln t)^{\hat{\nu}_u}} \right),
\end{equation}
where $\zeta_x=1$, $\hat{\zeta}_x=1/2$, $\nu_x=1/2$, $\hat{\nu}_x=1/4$, $\zeta_m=0$, $\hat{\zeta}_m=1/2$, $\nu_m=0$, and $\hat{\nu}_m=1/2$. The exponents for $\mathcal{T}$ are the same as those for $m$.

For $\beta'>\mathcal{B}_{\text{2D}}$, the system transitions to a unique subdiffusive universality class with $\alpha=9/10$ and $\hat{\alpha}=0$. These exponents stand out in the context of subdiffusion. The position distribution scaling exponents are $\zeta_x=1$ and $\nu_x=9/20$. These findings highlight the distinctive behaviors that emerge in two dimensions when there is a strong attraction to the other agent's debris.

\begin{table*}[b]
  \caption{\label{tab:2d}
  \textbf{The asymptotic behavior of various measures for 2D model systems.}
  All coefficients $a$ for each measure are independent and may depend on the values $\beta$ and $\beta'$. $D_0$ is the initial ($t=0$) distance between two agents. $\mathcal{B}_{\text{2D}}$ is a function of $\beta$.
  }
  \begin{tabular}{| c | c | c | c |}
    \hline
    & 
    $0 \le \beta' \le \beta$ & 
    $\beta' > \mathcal{B}_{\text{2D}}$ &
    $\beta' = \beta = 0$ 
    \\ \hline
    $R^2_{t\to \infty}$  & 
    $t (\ln t)^{1/2}$ & 
    $t^{9/10}$ &
    $t$
    \\ \hline
    $D^2_{t\to \infty} $ & 
    $R^2_{t\to\infty}$ & 
    $R^2_{t\to\infty}$ &
    $R^2_{t\to\infty}$ 
    \\ \hline
    $P(\vec{x},t) $ & 
    \makecell
    {
      $t^{-1} (\ln t)^{-1/2} f \left( \dfrac{\vec{x}}{t^{1/2} (\ln t)^{1/4} } \right)$
    } & 
    \makecell
    {
      $t^{-1} f \left( \dfrac{\vec{x}-\vec{D}_0}{t^{9/20}}\right)$
    } &
    \makecell
    {
      $t^{-1} f \left( \dfrac{\vec{x}}{t^{1/2}}\right)$ \\
      $f(z) = e^{-az^2}$
    }
    \\ \hline
    $P(m,t)$ & 
    \makecell
    {
      $(\ln t)^{-1/2} f\left(\dfrac{m}{(\ln t)^{1/2}} \right)$, \\ 
      $f(z) = e^{-az}$
    } & 
    not conclusive &
    \makecell
    {
      $(\ln t)^{-1} f \left( \dfrac{m}{\ln t} \right)$ \\
      $f(z) = e^{-az}$
    }
    \\ \hline
  \end{tabular}
\end{table*}

%==================================================
\section{Conclusion}

Our study introduces a minimal model for a class of systems comprising two distinct agents that avoid retracing their own paths while seeking each other. This coupled dynamic captures the interplay between self-avoidance and attraction to the other agent, the trade-off of which is controlled by two external parameters, $\beta$ and $\beta'$. The resulting dynamics reveal a rich landscape of emergent behaviors, including distinct diffusion phases---superdiffusive and subdiffusive---that arise from the competition between these effects.

A key outcome of this work is the identification of new universality classes for coupled random walks, characterized by distinct scaling exponents and probability distributions. Our phase diagram (see Fig.~\ref{fig:rws_phases}) summarizes these results: for $\beta' \leq \beta$, the model recovers the true self-avoiding walk (TSAW) behavior in both one and two dimensions. However, for $\beta' > \beta$, new universality classes emerge. In one dimension, we identify a subdiffusive regime with $\alpha = 1/2$ when $\beta = 0$ and $\beta' > 0$. For $\beta' > \mathcal{B}$, the system exhibits a regime with $\alpha \approx 1^-$, where the scaling exponent is slightly less than 1. While this exponent is close to that of a normal random walk ($\alpha=1$), the position distributions are fat-tailed and non-Gaussian, distinguishing this behavior as a separate universality class. In two dimensions, we uncover a novel universality class with $\alpha = 9/10$ for $\beta' > \mathcal{B}_{\text{2D}}$. To our knowledge, these coupled universality classes have not been previously reported and provide a novel perspective on how memory, feedback, and interaction shape emergent scaling behavior.

The emergence of fat-tailed position distributions in the subdiffusive regime suggests potential shifts in exploration strategies, favoring either highly localized search patterns or rare, long-range excursions. Such behaviors, arising from the interplay of memory and interaction, could have practical implications for systems where agent coupling guides collective dynamics, such as in ecological search processes, pheromone-based navigation in insect populations, or synthetic multi-agent systems. Understanding these dynamics may inform strategies for controlling encounter frequencies, optimizing search efficiency, or manipulating collective behavior in engineered systems.

Our results suggest that coupling strengths can be tuned to control agent behavior, potentially leading to applications in systems where localization or dispersion of agents is desired. For example, in biological contexts, the principles identified here may inform the design of pheromone-based strategies to concentrate or disperse populations. More broadly, our model provides a theoretical basis for manipulating encounter frequencies, optimizing search efficiency, or engineering localized trapping mechanisms in stochastic multi-agent systems.

By linking these findings to existing frameworks of anomalous diffusion, ecological search processes \cite{voituriez2011,chupeau2015}, and statistical physics \cite{newman1999}, we advance the theoretical understanding of transport phenomena in systems with memory and coupling. The identification of these scaling regimes and the role of feedback mechanisms highlight the broader relevance of our model beyond its biological inspiration, offering insights into a wide class of stochastic systems with interaction feedback.

While the model captures core aspects of coupled agent dynamics, several natural extensions present exciting directions for future research. Incorporating factors such as environmental heterogeneity, non-reciprocal interactions, or alternative transition rules (e.g., subexponential probabilities) could reveal even richer behaviors and scaling regimes. The framework developed here lays a foundation for exploring how memory, feedback, and interaction networks shape emergent dynamics in multi-agent systems. Beyond its immediate context, this model may also inspire new approaches in algorithmic search, collective behavior, and control strategies in stochastic systems.

%--------------------------------------

\begin{acknowledgments}
The authors thank the Australian Research Council for its support for this project through the Discovery program (FL190100080).
We acknowledge access to computational resources provided by the Pawsey Supercomputing Centre with funding from the Australian Government and the government of Western Australia, and the National Computational Infrastructure (NCI Australia), an NCRIS enabled capability supported by the Australian Government. We also acknowledge support of the Research Computing Centre at The University of Queensland.
\end{acknowledgments}
%==================================================
\bibliography{refs}
%==================================================

\newpage

\pagenumbering{gobble}
\renewcommand{\thepage}{S\arabic{page}}

\newcounter{x}
\forloop{x}{1}{\value{x} < 8}
{
	
	\centerline{ \includegraphics[width=1.2\textwidth,page=\value{x}]{suppl.pdf}}
}
\end{document}